\documentclass[12pt]{article}
\usepackage{graphicx}
\usepackage{epsfig}
\usepackage{epstopdf}
\DeclareGraphicsExtensions{.pdf,.eps,.png,.jpg,.mps}

\def\lsim{\mathrel{\rlap{\lower3pt\hbox{\hskip0pt$\sim$}}
     \raise1pt\hbox{$<$}}}         %less than or approx. symbol
\def\gsim{\mathrel{\rlap{\lower4pt\hbox{\hskip1pt$\sim$}}
     \raise1pt\hbox{$>$}}}         %greater than or approx. symbol

\usepackage{amsmath}
\usepackage{amsfonts}

\begin{document}
\begin{titlepage}

\centerline{\Large \bf Can Turnover Go to Zero?}
\medskip

\centerline{Zura Kakushadze$^\S$$^\dag$\footnote{\, \tt Email: zura@quantigic.com}}

\bigskip

\centerline{\em $^\S$ Quantigic$^\circledR$ Solutions LLC}
\centerline{\em 1127 High Ridge Road \#135, Stamford, CT 06905\,\,\footnote{\, DISCLAIMER: This address is used by the corresponding author for no
purpose other than to indicate his professional affiliation as is customary in
publications. In particular, the contents of this paper
are not intended as an investment, legal, tax or any other such advice,
and in no way represent views of Quantigic® Solutions LLC,
the website \underline{www.quantigic.com} or any of their other affiliates.
}}
\centerline{\em $^\dag$ Department of Physics, University of Connecticut}
\centerline{\em 1 University Place, Stamford, CT 06901}
\medskip
\centerline{(May 30, 2014; revised: July 14, 2014)}

\bigskip
\medskip

\begin{abstract}
{}Internal crossing of trades between multiple alpha streams results in portfolio turnover reduction.
Turnover reduction can be modeled using the correlation structure of the alpha streams. As more and more alphas are added, generally
turnover reduces. In this note we use a factor model approach to address the question of whether the turnover goes to zero or a finite limit as
the number of alphas $N$ goes to infinity. We argue that the limiting turnover value is determined by the number of alpha clusters $F$,
not the number of alphas $N$. This limiting value behaves according to the ``power law" $\sim F^{-3/2}$. So, to achieve zero limiting turnover, the number of alpha clusters must go to infinity along with the number of alphas. We further argue on general grounds that, if the number of underlying tradable instruments is finite, then the turnover cannot go to zero, which implies that the number of alpha clusters also appears to be finite.
\end{abstract}

\bigskip
\medskip

\end{titlepage}

\newpage

\section{Introduction and Summary}

{}When multiple alpha streams are traded on the same hedge fund platform, if the execution platform allows this, it makes sense to cross trades between different alpha streams thereby saving on transaction costs.\footnote{\, An illustrative discussion of internal crossing and its benefits can be found in (Kakushadze and Liew, 2014). A spectral model of turnover reduction, which we discuss in this note, was recently proposed in (Kakushadze, 2014a). For a partial list of hedge fund literature, see, {\em e.g.}, (Ackerman {\em et al}, 1999), (Agarwal and Naik, 2000a, 2000b), (Amin and Kat, 2003), (Asness {\em et al}, 2001), (Brooks and Kat, 2002), (Brown {\em et al}, 1999), (Chan {\em et al}, 2006), (Edwards and Caglayan, 2001), (Edwards and Liew, 1999a, 1999b), (Fung and Hsieh, 1999, 2000, 2001), (Kao, 2002), (Liang, 1999, 2000, 2001), (Lo, 2001), (Schneeweis {\em et al}, 1996).} When internal crossing is employed, portfolio turnover is reduced.

{}As more and more alphas are added, generally the percentage of the dollar turnover with respect to the total dollar investment -- which percentage we refer to simply as ``turnover" -- is expected to decrease. The more correlated the trades are, the more correlated the alphas are, and the more correlated the trades are, the lower the internal crossing is expected to be. Therefore, while turnover reduction is not necessarily a simple ({\em e.g.}, linear) function of alpha correlations, it is clear that it is somehow related to them, so one can try to model turnover reduction based on alpha correlations. In (Kakushadze, 2014a) a spectral model of turnover reduction based on the alpha correlation structure was proposed using {\em principal component} analysis of the alpha correlation matrix. A simplified version of this model was used in (Kakushadze and Liew, 2014) to argue that when the number of alphas is large, the portfolio turnover has a non-vanishing limit.

{}In this note we use a factor model approach -- as {\em complementary} to the principal component approach of (Kakushadze, 2014a) -- to address the question of whether the turnover goes to zero or a finite limit as the number of alphas $N$ goes to infinity. We argue that the limiting turnover value is determined by the number of alpha clusters $F$, not the number of alphas $N$. This limiting value behaves according to the ``power law" $\sim F^{-3/2}$. So, to achieve zero limiting turnover, the number of alpha clusters must go to infinity along with the number of alphas. We further argue on general grounds that, if the number of underlying tradable instruments is finite, then the turnover cannot go to zero, which implies that the number of alpha clusters also appears to be finite. The large $N$ limit (Kakushadze and Liew, 2014), (Kakushadze, 2014a) plays an important simplifying role in our discussion.

{}The remainder of this paper is organized as follows. Definitions are in Section 2. We discuss a factor model approach for alpha streams in Section 3. In Section 4 we briefly review the spectral model of turnover reduction of (Kakushadze, 2014a). In Section 5 we use the factor model approach of Section 3 to study what happens to the turnover in the large $N$ limit, including the effects of the alpha clusters, specific risk, off-diagonal elements of the factor covariance matrix, and style and general non-binary risk factors on the turnover. In Section 6 we discuss a practical application of our results in Section 5. In Section 7 we discuss methods for estimating lower and upper bounds for the number of alpha clusters. In Section 8 we further argue -- this is a non-rigorous ``theorem" of sorts -- on general grounds that, if the number of underlying tradable instruments is finite, then the turnover cannot go to zero, which implies that the number of alpha clusters also appears to be finite. Our main result is given by Eqns. (\ref{power.law}) and (\ref{power.law.1}), and the ``theorem" in Section \ref{theorem}.

\section{Definitions}\label{Definitions}

{}We have $N$ alphas $\alpha_i$, $i=1,\dots,N$. Each alpha is actually a time series $\alpha_i(t_s)$, $s=0,1,\dots,M$, where $t_0$ is the most recent time. Below $\alpha_i$ refers to $\alpha_i(t_0)$.

{}Let $C_{ij}$ be the covariance matrix of the $N$ time series $\alpha_i(t_s)$. Let $\Psi_{ij}$ be the corresponding correlation matrix, {\em i.e.},
\begin{equation}
 C_{ij} = \sigma_i~\sigma_j~\Psi_{ij}
\end{equation}
where $\Psi_{ii} = 1$.

{}In the papers (Kakushadze and Liew, 2014) and (Kakushadze, 2014a) it was argued that in the large $N$ limit the turnover does not necessarily go to zero and can go to a finite limit depending on the structure of the correlation matrix.\footnote{\, Caveats and limitations of modeling internal crossing and turnover reduction using the correlation matrix were discussed in (Kakushadze and Liew, 2014) and (Kakushadze, 2014a), so we will not repeat them here.} Our goal in this paper is to address the following question: What would make the turnover go to zero in the large $N$ limit?

\section{Factor Model}

{}Generally, the covariance matrix $C_{ij}$ can have the following undesirable properties. First, it can be (nearly) degenerate. Second, it may not be positive (semi-)definite.

{}Near degeneracy is caused by alphas that are almost 100\% correlated or anti-correlated and can be cured by simply removing such ``redundant" alphas.\footnote{\, See (Kakushadze, 2014a) for a more detailed discussion.} However, in practice, near degeneracy is usually caused by the fact that $M < N$ (in fact, in most practical applications $M \ll N$) and only $M$ eigenvalues of $C_{ij}$ are non-zero, while the remainder have ``small" values, which can be positive or negative. These small values are zeros distorted by computational rounding.\footnote{\, Actually, this assumes that there are no N/As in any of the alpha time series. If some or all alpha time series contain N/As in  non-uniform manner and the correlation matrix is computed by omitting such pair-wise N/As, then the resulting correlation matrix may have negative eigenvalues that are not ``small" in the sense used above, {\em i.e.}, they are not zeros distorted by computational rounding. The deformation method mentioned below can be applied in this case as well.\label{corrNAs}} In such cases, the solution is not to remove any alphas (as they are not necessarily ``redundant"), but to deform the covariance matrix so it is positive-definite. One such deformation based on (Rebonato and J\"ackel, 1999) was discussed in (Kakushadze, 2014a) (see Subsection 3.1 thereof). Another, perhaps more commonly used method,\footnote{\, More commonly used in the case of stocks, that is.} is to employ a factor model approach.

{}The aforementioned issue arising when $M < N$ conceptually is the same as the problem of modeling risk of a portfolio consisting of a large number $N_S$ of stocks.\footnote{\, The high-level math formalism is the same, albeit details are different (Kakushadze, 2014b).} Unless the number $M_S + 1$ of observations in the corresponding time series is large compared with the number of stocks, the covariance matrix based on such time series is not expected to be very stable. Furthermore, if $M_S < N_S$, as above, the correlation matrix is degenerate. One way to circumvent this problem in the case of stocks is that one builds a factor model, where instead of $N_S$ stocks one deals with $F_S$ risk factors, where $F_S\ll N_S$ and $F_S\lsim M_S$.

{}The same can be done with alphas. Instead of $N$ alphas, one deals with $F\ll N$ risk factors and the covariance matrix $C_{ij}$ is replaced by $\Gamma_{ij}$ given by
\begin{eqnarray}\label{Gamma}
 &&\Gamma \equiv \Xi + \Omega~\Phi~\Omega^T\\
 && \Xi_{ij} \equiv \xi_i^2 ~\delta_{ij}
\end{eqnarray}
where $\xi_i$ is the specific risk for each $\alpha_i$; $\Omega_{iA}$ is an $N\times F$ factor loadings matrix; and $\Phi_{AB}$ is the factor covariance matrix, $A,B=1,\dots,F$. {\em I.e.}, the random processes $\Upsilon_i$ corresponding to $N$ alphas are modeled via $N$ random processes $z_i$ (corresponding to specific risk) together with $F$ random processes $f_A$ (corresponding to factor risk):
\begin{eqnarray}
 &&\Upsilon_i = z_i + \sum_{A=1}^F \Omega_{iA}~f_A\\
 &&\left<z_i, z_j\right> = \Xi_{ij}\\
 &&\left<z_i, f_A\right> = 0\\
 &&\left<f_A, f_B\right> = \Phi_{AB}\\
 &&\left<\Upsilon_i, \Upsilon_j\right> = \Gamma_{ij}
\end{eqnarray}
Instead of an $N \times N$ covariance matrix $C_{ij}$ we now have an $F \times F$ covariance matrix $\Phi_{AB}$, which is much more stable than $C_{ij}$.

{}One approach to constructing a factor model for alphas is to have $F_{style}$ style risk factors and $F_{cluster}$ cluster risk factors. In the case of stocks, cluster risk factors are usually referred to as industry risk factors. Since here we are dealing with alphas, we will refer to such risk factors as cluster risk factors. Generally, clusters can be thought of as groupings of alphas based on some similarity criteria, which to a large extent boil down to how closely alphas are correlated with each other, albeit such similarity criteria need not be (but can be) based on sample correlation matrix -- see below. On the other hand, style risk factors are not based on any similarity criteria; instead, they are based on some estimated (or measured) properties of alphas -- thus, in a given cluster we can have alphas with vastly different values of a given style factor. In the case of alphas, the following style factors {\em a priori} appear possible, at least when the underlying tradables are stocks: 1) volatility, 2) turnover, and 3) momentum. Another (perhaps more difficult to implement) style factor one may wish to consider is capacity, {\em i.e.}, how much capital each alpha can absorb. One may also wish to add other style factors depending on how alphas are constructed, {\em etc.}\footnote{\, For a discussion of a general framework for alpha factor models, see (Kakushadze, 2014b).}

{}In the case of stocks, cluster factors are (usually) based on industry classification. In the case of alphas, one can use a taxonomy of alphas, {\em i.e.}, one classifies alphas according to how they are constructed -- if the required data is available, that is. Out of thousands of alphas one may construct, many are very similar to each other by construction. It is then clear that this similarity makes them more correlated, just as stocks belonging to the same industry are more correlated.\footnote{\, Conversely, if it were the case that $M\gg N$, high (anti-)correlations of alphas in the sample correlation matrix would imply their similarity. However, in practice we have $M < N$ and, in fact, $M\ll N$, so deducing similarity from the sample correlation matrix is not as straightforward. There exist (typically, proprietary) algorithms for building binary clusters based on the sample correlation matrix and/or its principal components. Such clusters generally are expected to inherit, at least to a large extent, the usual out-of-sample instability of the off-diagonal elements of the sample correlation matrix. However, in cases where the detailed information about how alphas are constructed is not available -- {\em e.g.}, the only data available could be the position data, {\em i.e.}, vectors of desired holdings to achieve by some times $T_1, T_2, \dots$ -- then such algorithms might come handy.\label{MN}} Just as in the case of stocks, it therefore makes sense to treat clusters as risk factors and model correlations between alphas based on such risk factors as opposed to computing them directly based on a large number $N$ of the time series corresponding to individual alphas. Note that a bonus of the factor model approach is that the matrix $\Gamma_{ij}$ is automatically positive-definite if all $\xi_i$ are non-zero and the factor covariance matrix is positive-definite (albeit this condition can be relaxed).

\section{Spectral Model}

{}In the paper (Kakushadze, 2014a) we discussed a spectral model of turnover reduction, which suggests that the behavior of the turnover in the large $N$ limit can be approximated as follows:\footnote{\, Let us note that, in the paper (Kakushadze, 2014a), Eq. (\ref{rho}) below was obtained by observing that in the context of the spectral model the largest eigenvalue of $\Psi_{ij}$ has the leading contribution to $T$ in the large $N$ limit. If other contributions are not suppressed, then (\ref{rho}) can be thought of as an approximate lower bound for turnover reduction.}
\begin{equation}\label{T}
 T \approx \rho_* ~\sum_{i=1}^N \tau_i~\left|w_i\right|
\end{equation}
where (the modulus $|\cdot|$ stands for the absolute value of the sum)
\begin{equation}\label{rho}
 \rho_* \equiv {\psi^{(1)}\over{N\sqrt{N}}}~ \left|\sum_{i=1}^N V^{(1)}_{i}\right|
\end{equation}
and $\psi^{(1)}$ is the largest eigenvalue of the correlation matrix $\Psi_{ij}$, while $V^{(1)}$ is the corresponding right eigenvector of $\Psi_{ij}$ normalized such that\footnote{\, Here the basis of $\alpha_i$ is taken ({\em i.e.}, the signs of $\alpha_i$ are chosen) such that $\sum_{i,j=1}^N\Psi_{ij} \equiv N^2\rho^\prime = N(1 + (N-1){\overline \rho})$ is maximized (${\overline \rho}$ is the mean correlation). This is because in the zeroth approximation $\rho_*$ is given by $\rho^\prime$ (Kakushadze, 2014a). {\em E.g.}, if we have only two alphas $\alpha_1$ and $\alpha_2$ and $\Psi_{12} < 0$, we can take $\alpha_1^\prime \equiv \alpha_1$ and $\alpha^\prime_2\equiv -\alpha_2$, so the resulting correlation $\Psi^\prime_{12} > 0$.\label{rho.prime}}
\begin{equation}
 \sum_{i=1}^N \left(V^{(1)}_i \right)^2 = 1
\end{equation}
Also, $w_i$ are the weights with which alphas are combined, $\sum_{i=1}^N \left|w_i\right| = 1$, and $\tau_i$ are the turnovers corresponding to individual alphas $\alpha_i$.

{}For illustrative purposes, let us consider the case where all off-diagonal elements of the correlation matrix $\Psi_{ij}$ are identical: $\Psi_{ij} = \rho$ ($i\not=j$). Also, let all $T_i \equiv \tau_i~\left|w_i\right|$ be identical, let $\psi^{(p)}$ be the eigenvalues of $\Psi_{ij}$, and let $V^{(p)}$ be the corresponding eigenvectors. Then, in the basis where $\psi^{(1)}$ is the largest eigenvalue, we have (note that $V^{(1)}_{i} \equiv 1/\sqrt{N}$, $i=1,\dots,N$ in this case):
\begin{eqnarray}
 &&\sum_{i=1}^N V^{(1)}_{i}~T_i = {1\over\sqrt{N}}~\sum_{i=1}^N T_i\\
 &&\sum_{i=1}^N V^{(p)}_{i}~T_i = 0,~~~p>1\\
 &&\psi^{(1)}=1+(N-1)~\rho\\
 &&\psi^{(p)} = 1- \rho,~~~p>1\\
 &&\rho_* = {{1+(N-1)~\rho}\over N}
\end{eqnarray}
which reproduces Eq. (12) in (Kakushadze and Liew, 2014). The first two equations above capture the essence of why the spectral model of (Kakushadze, 2014a) is expected to be a good approximation in the large $N$ limit even in the case of a general correlation matrix: for generic configurations of $T_i$ ({\em i.e.}, such that $T_i$ are not highly skewed and are reasonably distributed around their mean value), the contribution to $\rho_*$ due to the first principal component (corresponding to the largest eigenvalue) is dominant,\footnote{\, Another way of thinking about this is that the first principal component $V^{(1)}$ is closest to the full $SO(N)$ rotational invariance, {\em i.e.}, to the normalized unit vector ${\overline V}_i\equiv 1/\sqrt{N}$, $i=1,\dots. N$.\label{SO(N)}} while contributions due to other principal components are suppressed by powers of $1/N$. In the above case of uniform correlations, let us consider two extreme values of $\rho$. First, when $\rho = 1$, {\em i.e.}, when all alphas are 100\% correlated with each other, we have $\psi^{(1)} = N$ and $\rho_* = 1$, and there is no turnover reduction. On the other hand, when $\rho = 0$, {\em i.e.}, when none of the alphas are correlated with each other, we have $\psi^{(1)} = 1$ and $\rho_* = 1/N$, and the turnover goes to zero in the large $N$ limit. The above extreme cases are in agreement with what we expect based on an intuitive picture of turnover reduction (Kakushadze and Liew, 2014).

\section{Factor Model and Turnover}

{}In this section our goal is to study turnover reduction in the large $N$ limit using the factor model, which provides a simple computational framework for understanding the behavior of the turnover reduction coefficient $\rho_*$ with increasing $N$. The spectral model formula (\ref{rho}), obtained using the principal component analysis (Kakushadze, 2014a), provides a well-defined prescription for estimating the turnover reduction coefficient $\rho_*$ for a given alpha correlation matrix.\footnote{\, Even when $M < N$, {\em i.e.}, when the sample correlation matrix is singular, (\ref{rho}) is still applicable. This is because it uses the first principal component $V^{(1)}$ and the largest eigenvalue $\psi^{(1)}$, whose dependence on $M$ does not alter the results dramatically. One way to see this is to deform a singular correlation matrix using the method discussed in Subsection 3.1 of (Kakushadze, 2014a) based on (Rebonato and J\"ackel, 1999) with the non-positive eigenvalues replaced by the smallest positive eigenvalue, as in Section 7. Thus, for the hedge fund data we use in Section 7, for the largest eigenvalue the difference between the undeformed and deformed cases is only about 24\%.} However, our goal here is to gain {\em intuitive} insight into the large $N$ behavior of $\rho_*$. The factor model approach provides a convenient calculational playground is this regard: we have $F$ risk factors (specified via the $N\times F$ factor loadings matrix $\Omega_{iA}$), the $F\times F$ factor covariance matrix $\Phi_{AB}$, and the diagonal $N\times N$ specific risk matrix $\Xi_{ij}$. For instance, we can ask: what is the dependence of $\rho_*$ on $F$ in the large $N$ limit?

{}We will start with a simplified factor model\footnote{\, We add complexity below, including {\em non-binary} factor loadings -- the simplified factor model serves the purpose of illustrating the key issues without overcomplicating them with unnecessary math. Here we choose binary factor loadings for calculational convenience. As we will see below, the simplified factor model captures the key features of the large $N$ behavior. Also, note that the simplified model of (Kakushadze and Liew, 2014) is a special case of a general one-factor model.} where i) specific risks $\xi_i$ are set to zero, ii) there are no style risk factors, and all cluster risk factors are ``binary" in the sense that each $\alpha_i$ belongs to one and only one cluster, and iii) the factor covariance matrix $\Phi_{AB}$ is diagonal:\footnote{\, Here the values of the factor loadings elements need not be ``binary". If $\Omega_{iA} = \omega_{iA}~\delta_{G(i), A}$ with non-binary $\omega_{iA}$, the result is unchanged, both here and in Subsection 5.2 (non-diagonal factor covariance matrix). What is important here is the binary membership of alphas in clusters.}
\begin{eqnarray}
 &&\xi_i \equiv 0\\
 &&\Omega_{iA} = \delta_{G(i), A}\\
 &&\Phi_{AB} = \phi_A~\delta_{AB}
\end{eqnarray}
where
\begin{equation}
 G:\{1,\dots, N\} \mapsto \{1,\dots, F\}
\end{equation}
is the map between alphas and clusters.

{}We have:
\begin{eqnarray}
 && \Gamma_{ij} = \phi_{G(i)}~\delta_{G(i), G(j)}\\
 && \sigma_i^2 \equiv \Gamma_{ii} = \phi_{G(i)}\\
 && \Psi_{ij} \equiv {1\over{\sigma_i~\sigma_j}}~\Gamma_{ij} = \delta_{G(i), G(j)}
\end{eqnarray}
and the correlation matrix is block-diagonal with $F$ blocks corresponding to $F$ clusters, and each diagonal block has all elements equal 1.

{}The correlation matrix $\Psi_{ij}$ has $(N - F)$ null eigenvalues\footnote{\, There is no reason for alarm -- these eigenvalues are null because we set the specific risks to zero -- see below.} and $F$ eigenvalues equal to:
\begin{equation}
 \psi^{(A)} = N_A
\end{equation}
where
\begin{equation}
 N_A \equiv \sum_{i=1}^N \delta_{G(i), A}
\end{equation}
is the number of alphas that belong to the cluster labeled by $A$. Note that
\begin{equation}
 \sum_{A=1}^F N_A = N
\end{equation}
Also, the right eigenvectors $V^{(A)}_i$ corresponding to the eigenvalues $\psi^{(A)}$ are given by
\begin{equation}\label{eigenvectors}
 V^{(A)}_i = {1\over \sqrt{N_A}}~\delta_{G(i), A}
\end{equation}
and are normalized such that
\begin{equation}
 \sum_{i=1}^N \left(V^{(A)}_i \right)^2 = 1
\end{equation}
Note that
\begin{equation}\label{power.law}
 \rho_* = \left({N_* \over N}\right)^{3\over 2} \gsim {1\over F^{3\over 2}}
\end{equation}
where
\begin{equation}
 N_* \equiv \mbox{max}\left(N_A, A=1,\dots,F\right)
\end{equation}
Therefore, for fixed $F$ simply increasing $N$ does not reduce turnover indefinitely.

{}This is because adding more and more alphas that belong to the same clusters does not diversify them as alphas in the same clusters are correlated with each other. To reduce $\rho_*$ one also needs to increase $F$, {\em i.e.}, to add alphas that belong to new clusters. Also, note that if $F$ is fixed, adding more and more alphas can decrease $\rho_*$ (unless all alphas are added to the cluster with the largest $N_A$), but it does not decrease it to zero.

{}In this regard, let us consider the case where $N\rightarrow\infty$ with $F = \mbox{fixed}$. From (\ref{power.law}) it follows that $\rho_*$ is minimized when $N_*$ is minimized, where $N_*$ is the largest $N_A$. It then follows that $N_* = \mbox{ceiling}(N / F)$, where $\mbox{ceiling}(x)\equiv \lceil x \rceil$ refers to the smallest integer not less than $x$, and similarly $\mbox{floor}(x)\equiv \lfloor x \rfloor$ refers to the largest integer not greater than $x$. Consequently, in the $N\rightarrow \infty$ limit the minimum of $\rho_*$ is at $(\rho_*)_{\rm{\scriptstyle{min}}} \equiv F^{-3/2}$. For any finite $N$ the corresponding distribution has $F_-$ values of $N_A = \mbox{floor}(N / F)$ and $F_+$ values of $N_A = \mbox{ceiling}(N / F)$, where $F_- \equiv F - F_+$ and $F_+ \equiv N - F~\mbox{floor}(N / F)$. Note that $0 \leq F_+ < F$, and in the case where $F_+ = 0$ we have $\mbox{ceiling}(N / F) = \mbox{floor}(N / F) = N/F$.

\subsection{Effect of Specific Risk}

{}Above we assumed zero specific risk. In this subsection we study effects of non-zero specific risk. To keep things simple, we will assume that
\begin{eqnarray}
 &&\xi_i = {\widetilde \xi}_{G(i)}\\
 &&\Omega_{iA} = \delta_{G(i), A}\\
 &&\Phi_{AB} = \phi_A~\delta_{AB}
\end{eqnarray}
where ${\widetilde \xi}_A$, $A=1,\dots,F$ are some specific risks corresponding to each cluster. {\em I.e.}, we assume that specific risk is uniform within each cluster, which corresponds to an approximation where specific risk within each cluster is replaced (be it directly or logarithmically), {\em e.g.}, by mean (or median) specific risk or square root of mean (median) specific variance for that cluster. We do this here for computational simplicity -- as we will see, the effect of specific risk on turnover reduction is subleading in the large $N$ limit, and so is the effect of nonuniform specific risk within each cluster.

{}With the above assumptions we now have:
\begin{eqnarray}
 && \Gamma_{ij} = {\widetilde \xi}^2_{G(i)}~\delta_{ij} + \phi_{G(i)}~\delta_{G(i), G(j)}\\
 && \sigma_i^2 \equiv \Gamma_{ii} = {\widetilde \xi}^2_{G(i)} + \phi_{G(i)}\\
 && \Psi_{ij} \equiv {1\over{\sigma_i~\sigma_j}}~\Gamma_{ij} = {{\widetilde \xi}^2_{G(i)} \over {{\widetilde \xi}^2_{G(i)} + \phi_{G(i)}}}~\delta_{ij} +
 {\phi_{G(i)} \over {{\widetilde \xi}^2_{G(i)} + \phi_{G(i)}}}~\delta_{G(i), G(j)}
\end{eqnarray}
This correlation matrix has the following eigenvalue structure. For each $A =1,\dots,F$, it has $(N_A - 1)$ eigenvalues equal
\begin{equation}
 {\widetilde \psi}^{(A)} = {{\widetilde \xi}^2_A \over {{\widetilde \xi}^2_A + \phi_A}},~~~{\widetilde d}_A = N_A - 1,~~~A=1,\dots,F
\end{equation}
where ${\widetilde d}_A$ is the degeneracy of each such eigenvalue. This gives total of $N - F$ eigenvalues. The remaining $F$ eigenvalues (each with unit degeneracy $d_A = 1$) are given by:
\begin{equation}
 \psi^{(A)} = {{{\widetilde \xi}^2_A + N_A~\phi_A} \over {{\widetilde \xi}^2_A + \phi_A}},~~~d_A = 1,~~~A=1,\dots,F
\end{equation}
This reduces to the previous result when all ${\widetilde \xi}_A \equiv 0$.

{}The right eigenvectors $V^{(A)}_i$ corresponding to the eigenvalues $\psi^{(A)}$, as before, are given by
\begin{equation}\label{eigenvectors1}
 V^{(A)}_i = {1\over \sqrt{N_A}}~\delta_{G(i), A}
\end{equation}
So we have:
\begin{equation}\label{spec.risk}
 \rho_* = {{1 + {1 \over N_*}~\zeta_*} \over {1 + \zeta_*}}~\left({N_* \over N}\right)^{3\over 2}
\end{equation}
where
\begin{equation}
 \zeta_A \equiv {{\widetilde \xi}^2_A \over \phi_A}
\end{equation}
and $\zeta_* \equiv \zeta_A$ for the value of $A$ for which $N_A = N_*$ (assuming for the sake of simplicity that $N_A$ are all unique). As we see, specific risk does not affect the large $N$ behavior of $\rho_*$; instead, it simply amounts to reducing the overall coefficient in $\rho_*$, but does not affect the conclusions we arrived at in the zero specific risk case.

{}In this regard, {\em assuming} it is kept finite (see below), specific risk does not qualitatively affect turnover reduction in the large $N$ limit, albeit it is important, {\em e.g.}, in the weight optimization even in the large $N$ limit. Furthermore, above we assumed that specific risk is uniform across each cluster, but relaxing this assumption does not change the above conclusions relating to turnover reduction in the large $N$ limit.

{}Here the following clarifying remark is in order. As mentioned above, specific risk does further reduce the overall turnover reduction coefficient. If we take $\zeta_* \rightarrow \infty$, {\em i.e.}, the factor risk is negligible compared with specific risk, then turnover goes to zero. Indeed, in this case we simply have $N$ uncorrelated alphas and in the large $N$ limit we have $\rho_* \sim \sqrt{N_*/N^3}\rightarrow 0$. In the above analysis, when discussing the behavior or $\rho_*$ in the large $N$ limit, we assume that $N\rightarrow \infty$ (and, consequently, $N_*\rightarrow \infty$) with $\zeta_* = \mbox{fixed}$. Then we have $\rho_*\sim (N_*/N)^{3/2} / (1 + \zeta_*) \rightarrow \mbox{const.}$

\subsection{Non-diagonal Factor Covariance Matrix}\label{non.diag.fac.cov}

{}Another simplifying assumption we made above was that the factor covariance matrix is diagonal. In this subsection we relax this assumption. Since we already know that specific risk does not affect the qualitative picture, we will set it to zero in order not to overcomplicate things, but we will make no assumption about the factor covariance matrix:
\begin{eqnarray}
 &&\xi_i \equiv 0\\
 &&\Omega_{iA} = \delta_{G(i), A}
\end{eqnarray}
We now have:
\begin{eqnarray}
 && \Gamma_{ij} = \Phi_{G(i), G(j)}\\
 && \sigma_i^2 \equiv \Gamma_{ii} = \Phi_{G(i), G(i)}\\
 && \Psi_{ij} \equiv {1\over{\sigma_i~\sigma_j}}~\Gamma_{ij} = {\widehat \Psi}_{G(i), G(j)}
\end{eqnarray}
where ${\widehat \Psi}_{AB}$ is the factor correlation matrix:
\begin{equation}
 \Phi_{AB} \equiv \sqrt{\Phi_{AA}}~\sqrt{\Phi_{BB}}~{\widehat \Psi}_{AB}
\end{equation}
Then the generalization of the eigenvectors (\ref{eigenvectors}) for non-diagonal $\Phi_{AB}$ can be found as follows.

{}There are $F$ such eigenvectors. They are of the form
\begin{equation}
 V^{(A)}_i = {1\over \sqrt{N_{G(i)}}}~\chi^{(A)}_{G(i)}
\end{equation}
Note that
\begin{equation}
 \sum_{j = 1}^N \Psi_{ij}~V^{(A)}_j = \sum_{j = 1}^N {\widehat \Psi}_{G(i), G(j)}~{1\over \sqrt{N_{G(j)}}}~\chi^{(A)}_{G(j)} = {1\over \sqrt{N_{G(i)}}}~\sum_{B=1}^F {\widehat \Psi}^\prime_{G(i),B}~\chi^{(A)}_{B}
\end{equation}
where
\begin{equation}
 {\widehat \Psi}^\prime_{AB} \equiv \sqrt{N_A}~{\widehat \Psi}_{AB}~\sqrt{N_B}
\end{equation}
Note that ${\widehat \Psi}^\prime$ is a symmetric matrix and $\chi^{(A)}_{B}$ are the eigenvectors of ${\widehat \Psi}^\prime$:
\begin{equation}\label{eigen}
 \sum_{B=1}^F {\widehat \Psi}^\prime_{CB}~\chi^{(A)}_{B} = {\widehat \psi}^{(A)}~\chi^{(A)}_{C}
\end{equation}
where ${\widehat \psi}^{(A)}$ are the eigenvalues of ${\widehat \Psi}^\prime$.

{}So we have
\begin{equation}
 \sum_{j = 1}^N \Psi_{ij}~V^{(A)}_j = {\widehat \psi}^{(A)}~V^{(A)}_i
\end{equation}
and
\begin{equation}\label{norm}
 1 = \sum_{i=1}^N \left(V^{(A)}_i \right)^2 = \sum_{B=1}^F \left(\chi^{(A)}_B \right)^2
\end{equation}
fixes the normalization of $\chi^{(A)}_{B}$.

{}Next, note that the diagonal elements ${\widehat \Psi}_{AA} \equiv 1$, and (in matrix notation)
\begin{equation}
 {\widehat \Psi}^\prime = Q~{\widehat \Psi}~Q
\end{equation}
where
\begin{equation}
 Q \equiv \mbox{diag}\left(\sqrt{N_A}\right)
\end{equation}
is a diagonal $F \times F$ matrix. This implies that, since ${\widehat \Psi}_{AB}$ is positive definite, so is ${\widehat \Psi}^\prime_{AB}$. Furthermore, we have:
\begin{equation}
 \mbox{Tr}\left({\widehat \Psi}^\prime\right) = \sum_{A=1}^F N_A = N
\end{equation}
This implies that the off-diagonal elements of ${\widehat \Psi}_{AB}$ have the following effect on the eigenvalues ${\widehat \psi}^{(A)}$. When the off-diagonal elements of ${\widehat \Psi}_{AB}$ are zero, these eigenvalues are equal $N_A$. When they are nonzero, these eigenvalues generally are different from $N_A$, but they are still positive and their sum is still equal $N$:
\begin{eqnarray}
 && \forall A=1,\dots,F:~~~{\widehat \psi}^{(A)} > 0\\
 && \sum_{A=1}^F {\widehat \psi}^{(A)} = N
\end{eqnarray}
Also
\begin{equation}
 \left|\sum_{i=1}^N V^{(A)}_i \right| = \left|\sum_{B=1}^F \sqrt{N_B}~\chi^{(A)}_B\right|
\end{equation}
This implies that ($V^*_i$ is the eigenvector corresponding to ${\widehat \psi}^*$)
\begin{eqnarray}\label{maxpsi}
 && {\widehat \psi}^* \equiv \mbox{max}\left({\widehat \psi}^{(A)}, A=1,\dots,F\right)\gsim {N\over F}\\
 && \left|\sum_{i=1}^N V^*_i \right| \gsim \sqrt{N\over F}\label{sumV}
\end{eqnarray}
and the above conclusions relating to turnover reduction in the large $N$ limit are unchanged even if the factor covariance matrix $\Phi_{AB}$ is not diagonal -- the off-diagonal elements generally increase the lower bound. Appendix A discusses (\ref{sumV}) in more detail. Adding specific risk does not modify this result.

\subsection{Effect of Style Factors}

{}Adding style risk factors into the mix does not alter the above conclusions. {\em Assuming} there is a finite and small number of style risk factors,\footnote{\, This is not an unreasonable assumption. In the case of stocks, the number of style risk factors is substantially smaller than the number of industry risk factors. In the case of alphas, it appears that the number of basic style risk factors should be even smaller, albeit there is a trick to effectively increase their number -- see (Kakushadze, 2014b) for details.} their inclusion has subleading effect when the number of cluster risk factors is large.

{}A quick way to see this is to consider a simple factor model with zero specific risk and a single style factor $\Omega_i$. The $1\times 1$ factor covariance matrix can be absorbed into the definition of $\Omega_i$. We then have
\begin{eqnarray}
 &&\Gamma_{ij} = \Omega_i ~ \Omega_j\\
 &&\sigma^2_i \equiv \Gamma_{ii} = \Omega^2_i\\
 && \Psi_{ij} \equiv {1\over{\sigma_i~\sigma_j}}~\Gamma_{ij} = 1
\end{eqnarray}
This matrix has $(N-1)$ null eigenvalues and a single nonzero eigenvalue equal $N$. Adding a few more style factors can reduce the maximum eigenvalue of the correlation matrix only by a factor of order 1. It takes a large number of risk factors, be it binary or otherwise, to reduce the maximum eigenvalue of the correlation matrix substantially, which is what we found above.

\subsection{Non-binary Factor Loadings}

{}Above, for the sake of simplicity, we assumed that the factor loadings $\Omega_{iA}$ are binary. In this subsection we consider the case of arbitrary non-binary factor loadings. In this case it is convenient to absorb the factor covariance matrix into the definition of the factor loadings:
\begin{eqnarray}
 &&\Gamma = \Xi + {\widetilde \Omega}~{\widetilde \Omega}^T\\
 &&{\widetilde \Omega} \equiv \Omega~{\widetilde \Phi}\\
 &&{\widetilde \Phi}~{\widetilde\Phi}^T = \Phi
\end{eqnarray}
where ${\widetilde \Phi}_{AB}$ is the Cholesky decomposition of $\Phi_{AB}$, which is assumed to be positive-definite.

{}For the sake of simplicity, as we did before Subsection 5.1, let us set specific risk to zero -- we comment on the effect of specific risk below. Then the covariance matrix reads:
\begin{eqnarray}
 &&\Gamma_{ij} = \sum_{A=1}^F {\widetilde \Omega}_{iA}~{\widetilde \Omega}_{jA}\\
 &&\sigma^2_i \equiv\Gamma_{ii} = \sum_{A=1}^F {\widetilde \Omega}^2_{iA}\\
 &&\Psi_{ij} \equiv {1\over {\sigma_i~\sigma}_j}~\Gamma_{ij} = \sum_{A=1}^F \Lambda_{iA}~\Lambda_{jA}
\end{eqnarray}
where
\begin{equation}
 \Lambda_{iA} \equiv {1\over \sigma_i}~{\widetilde\Omega}_{iA}
\end{equation}
The correlation matrix $\Psi_{ij}$ has $N-F$ null eigenvalues and $F$ non-zero eigenvalues.\footnote{\, More precisely, these $F$ eigenvalues are nonzero assuming that the matrix $Q_{AB}$ defined below is nonsingular -- see below.} The latter are given by the eigenvalues $\psi^{(A)}$ of the matrix
\begin{equation}
 Q_{AB} \equiv \sum_{i=1}^N \Lambda_{iA}~\Lambda_{iB}
\end{equation}
whose properties also determine the corresponding eigenvectors of $\Psi_{ij}$.

{}Let
\begin{eqnarray}
 &&Q = W~Z~W^T\\
 &&Z = \mbox{diag}\left(\psi^{(A)}\right)
\end{eqnarray}
where $W$ is the $F\times F$ matrix, normalized such that $W^T~W = 1$, whose columns are the right eigenvectors of $Q_{AB}$. We can assume that all $\psi^{(A)}>0$, $A=1,\dots, F$ --  the matrix $Q_{AB}$ is positive-definite provided that the $N$-vectors $y_i^{(A)} \equiv \Lambda_{iA}$, $A=1,\dots,F$ are linearly independent, which we assume to be the case without loss of generality. Further, let
\begin{equation}
 V^{(A)}_i \equiv {1\over\sqrt{\psi^{(A)}}}~\sum_{B=1}^F \Lambda_{iB}~W_{BA}
\end{equation}
Then we have
\begin{eqnarray}
 &&\sum_{j=1}^N \Psi_{ij}~V^{(A)}_j = \psi^{(A)}~V^{(A)}_i\\
 &&\sum_{i=1}^N V^{(A)}_i~V^{(B)}_i = \delta_{AB}
\end{eqnarray}
So, $V^{(A)}_i$ are the properly normalized $F$ eigenvectors of $\Psi_{ij}$ corresponding to the eigenvalues $\psi^{(A)}$, which have the following properties:
\begin{eqnarray}
 &&\sum_{A=1}^N \psi^{(A)} = \mbox{Tr}(Q) = \mbox{Tr}(\Psi) = N\\
 \label{maxpsi.1}
 && \psi^* \equiv \mbox{max}\left({\widehat \psi}^{(A)}, A=1,\dots,F\right)\gsim {N\over F}
\end{eqnarray}
We still need to estimate
\begin{equation}\label{Vstar.1}
 \left|\sum_{i=1}^N V^*_i \right|
\end{equation}
where $V^*_i$ is the eigenvector corresponding to the largest eigenvalue $\psi^*$. Details are relegated to Appendix B -- thus, in the case of non-binary factor loadings we also find a finite limit for the turnover reduction coefficient in the large $N$ limit. Adding specific risk does not modify this result.

\section{Practical Application}

{}As we saw above, in the context of a factor model for alphas, turnover reduction does not necessarily go to zero by adding more and more alphas. One needs to add more and more {\em different} types of new alphas to achieve indefinite turnover reduction. In the factor model context this translates into adding alphas that form new clusters, not adding alphas into old clusters. Intuitively, this result is not surprising -- to achieve turnover reduction, one needs to add more and more alphas that are (almost) uncorrelated with the existing alphas. However, the factor model framework provides a way of quantifying this statement. In particular, it allows to design a simple test of whether new alphas have a potential for turnover reduction.

{}Thus, suppose we have $N$ alphas $\alpha_i$ that form $F$ clusters with the factor loadings $\Omega_{iA}$, $i=1,\dots,N$, $A=1,\dots,F$. Suppose new alphas are developed and we need to assess whether they have a potential for turnover reduction. The question is whether these new alphas form a new cluster.\footnote{\, As mentioned in Subsection 5.1, specific risk decreases turnover. If new alphas belong to an existing cluster but have higher specific risk, turnover could be reduced. However, specific risk is computed in the risk factor model {\em after} identifying all relevant risk factors, including any clusters.} For simplicity let us consider the case where the claim is that the new alphas are all such that they form a single new cluster. Let the old plus new alphas be $\alpha^\prime_{i^\prime}$, $i^\prime = 1,\dots,N^\prime$, where $N^\prime$ is the total number of the old plus new alphas and $(N^\prime - N)$ is the number of the new alphas. Let the new factor loadings matrix be $\Omega^\prime_{i^\prime A^\prime}$, $A^\prime = 1,\dots, F^\prime$, where $F^\prime = F+1$.

{}Now we can run two regressions (without intercept), first $\alpha_i$ over $\Omega_{iA}$, and second $\alpha^\prime_{i^\prime}$ over $\Omega^\prime_{i^\prime A^\prime}$. In R notations:
\begin{eqnarray}
 &&\alpha \sim -1 + \Omega\\
 &&\alpha^\prime \sim -1 + \Omega^\prime
\end{eqnarray}
One can now compare the F-statistic for each of these regressions. In actuality, $\alpha_i$ and $\alpha^\prime_{i^\prime}$ are time series: $\alpha_i(t_s)$ and $\alpha^\prime_{i^\prime}(t_s)$, $s=0,1,\dots,M$. So one can look, {\em e.g.}, at two time-series vectors of F-statistic and assess whether the hypothesis that new alphas form a new cluster has better overall $F$-statistic.\footnote{\, In comparing the two F-statistic vectors, one can remove (or smooth via standard techniques, {\em e.g.}, Winsorization) outliers to improve statistical significance of the comparison.}

{}The above discussion assumes that the factor loadings $\Omega_{iA}$ corresponding to the existing clusters are known. The question is, how does one identify these existing clusters in the first place? A somewhat ``primitive", albeit perhaps practical, way of building out a binary alpha classification is to use the method of this section recursively. If there are many independent alpha sources (developers), in the zeroth approximation they effectively label the clusters. Every time new alphas are developed, one needs to filter out those too highly correlated with the existing alphas. If one has a binary classification with $K$ clusters, then one can build a classification with $K+1$ clusters, and repeat this recursively. Most care is needed when $K$ is low -- once $K$ is large, there is more statistics available for filtering out redundant new alphas. To improve this method at low $K$, one can supplement it with algorithms mentioned in footnote \ref{MN}, where binary factor loadings are obtained using sample correlation matrix and/or principal components.\footnote{\, Such (again, typically proprietary) algorithms are usually based on ranking techniques. They are expected to work better precisely when $K$ is low, as there is a practical bound on $M$.}

\section{``Power Law" and Number of Clusters}\label{cluster.num}

{}Our main result above is that, using a factor model approach, we arrived at the ``power law" (\ref{power.law}), which suggests that the turnover reduction coefficient $\rho_*$ is controlled the number of clusters $F$:
\begin{equation}\label{power.law.1}
 \rho_* \gsim {1\over F^{3\over 2}}
\end{equation}
Here we ask two questions. First, can we estimate $F$ for a given correlation matrix $\Psi_{ij}$? Second, can we increase $F$ indefinitely by increasing $N$, or is there a limit?

{}To answer the first question, let us first recall from Eq. (\ref{maxpsi}) that
\begin{equation}
 \psi^* \gsim {N \over F}
\end{equation}
This gives us a simple method for estimating $F$:
\begin{equation}\label{lower}
 F\gsim {N \over \psi^*}
\end{equation}
Here the following remark is in order. The lower bound (\ref{lower}) is applicable assuming $N\lsim M$, so that the correlation matrix is nonsingular. If $M <N$, then the correlation matrix is singular and the number of clusters $F\leq M$, so depending on $M$ the lower bound (\ref{lower}) may or may not be informative. As an {\em illustrative}\footnote{\, We emphasize the adjective ``illustrative" for the reason that, because various hedge funds in this data do/did not all trade the same underlying instruments and also the corresponding time series are not 100\% overlapping (some hedge funds are dead, some are newer than others, {\em etc.}), it would not necessarily be correct to assume that their trades could be crossed. Therefore, we use this data only to {\em illustrate} various properties of the correlation matrix, and not necessarily to directly draw any conclusions about turnover reduction had these alpha streams actually crossed their trades.} example, let us use the same Morningstar data as in Fig.1 of (Kakushadze and Liew, 2014) and Fig.1 of (Kakushadze, 2014a), which is the data for 1990-2014 for $N = 657$ monthly hedge fund returns (HF). For the undeformed\footnote{\, The raw HF have non-uniform N/As, so the correlation matrix is computed by omitting such pair-wise N/As and, as mentioned in footnote \ref{corrNAs}, has negative eigenvalues, which are dealt with by deforming the correlation matrix using the method discussed in Subsection 3.1 of (Kakushadze, 2014a) based on (Rebonato and J\"ackel, 1999) with the non-positive eigenvalues replaced by the smallest positive eigenvalue (which is not a zero distorted by computational rounding) -- also see the end of Section 7.1. Here the values we give for $\psi^*$ and $F$ are rounded to 2 significant figures.\label{foot.RJ}} correlation matrix of the raw HF we have $\psi^*\approx 207$ and $F\gsim 3.17$, and for the corresponding deformed correlation matrix of the raw HF we have $\psi^*\approx 158$ and $F\gsim 4.15$. For the undeformed correlation matrix of the residuals (plus the intercepts, which have no effect) of the HF adjusted for RF (whose effect is small) and regressed over Mkt-RF and the Fama-French risk factors SMB, HML, WML, we have $\psi^*\approx 93.9$ and $F\gsim 7.00$, and for the corresponding deformed correlation matrix we have $\psi^*\approx 71.1$ and $F\gsim 9.24$. Note how regressing away Mkt-RF and the Fama-French risk factors, which evidently dominate in the raw HF, improves the lower bound. Also note that in this example $M$ is substantially larger than the lower bound we obtained, so the latter is informative.

\subsection{Regression Correlations}

{}Above we discussed how to obtain a lower bound for the number of clusters. A nontrivial method for estimating an upper bound for $F$ is as follows. Let $\Lambda_{iA}$, $A=1,\dots K$ be the $N \times K$ matrix of the first $K$ principal components of $\Psi_{ij}$, {\em i.e.}, the columns of $\Lambda_{iA}$ are the first $K$ principal components.\footnote{\, There are other ways of constructing $\Lambda_{iA}$, but for our purposes here building it out of the principal components suffices.} Next, let $\epsilon_i(t_s)$ be the time series of the regression residuals, which in R notations reads ({\em i.e.}, the regression is without intercept and with trivial weights):
\begin{equation}
 {\widetilde \alpha} \sim -1 + \Lambda
\end{equation}
where (see Section \ref{Definitions})
\begin{equation}
 {\widetilde \alpha}_i \equiv {\alpha_i \over \sigma_i}
\end{equation}
Note that
\begin{equation}
 \mbox{Cov}\left({\widetilde \alpha}_i,~{\widetilde \alpha}_j\right) = \mbox{Cor}\left(\alpha_i,~\alpha_j\right)= \Psi_{ij}
\end{equation}
In matrix notations we have:
\begin{equation}
 \epsilon(t_s) = \left(1 - Y\right)~{\widetilde \alpha}(t_s)
\end{equation}
where
\begin{equation}
 Y \equiv \Lambda~\left(\Lambda^T~\Lambda\right)^{-1}~\Lambda^T
\end{equation}
is a projection matrix: $Y^2 = Y$.

{}Note that:
\begin{equation}
 \mbox{Cov}\left({\epsilon}_i,~{\epsilon}_j\right) = \left[\left(1-Y\right)~\Psi~\left(1-Y\right)\right]_{ij} \equiv \Phi_{ij}
\end{equation}
Let
\begin{eqnarray}
 &&\xi_i^2 \equiv \Phi_{ii}\\
 &&{\widetilde \epsilon}_i\equiv {\epsilon_i\over\xi_i}
\end{eqnarray}
Then
\begin{equation}
 {\widetilde \Phi}_{ij}\equiv \mbox{Cov}\left({\widetilde \epsilon}_i,~{\widetilde \epsilon}_j\right) = \mbox{Cor}\left(\epsilon_i,~\epsilon_j\right) = {1\over\xi_i~\xi_j}~\Phi_{ij}
\end{equation}
{\em I.e.}, ${\widetilde \Phi}_{ij}$ is the correlation matrix of the regression residuals $\epsilon_i$. Moreover, to compute it, we do not need to know $\alpha_i$; we only need to know the correlation matrix $\Psi_{ij}$.

\begin{figure}
  \centerline{\epsfxsize 4.truein \epsfysize 4.truein\epsfbox{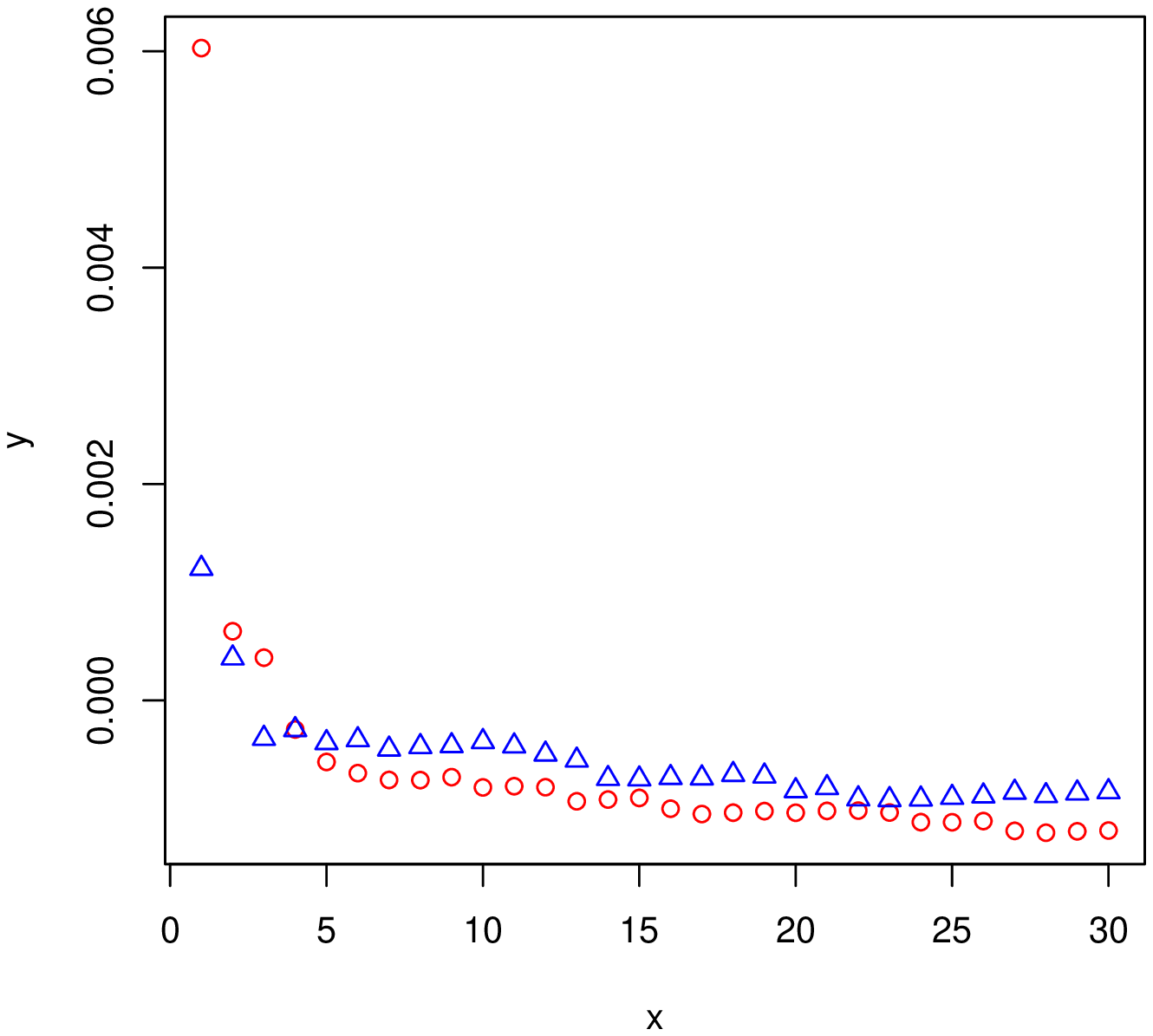}}
  \caption{$x$-axis: $K$; $y$-axis: $\zeta_1$. Circles: raw HF; triangles: HF regressed over Mkt-RF and Fama-French risk factors SMB, HML, WML. See Section \ref{cluster.num} for details.}\label{MeanCorr30}
\end{figure}

\begin{figure}
  \centerline{\epsfxsize 4.truein \epsfysize 4.truein\epsfbox{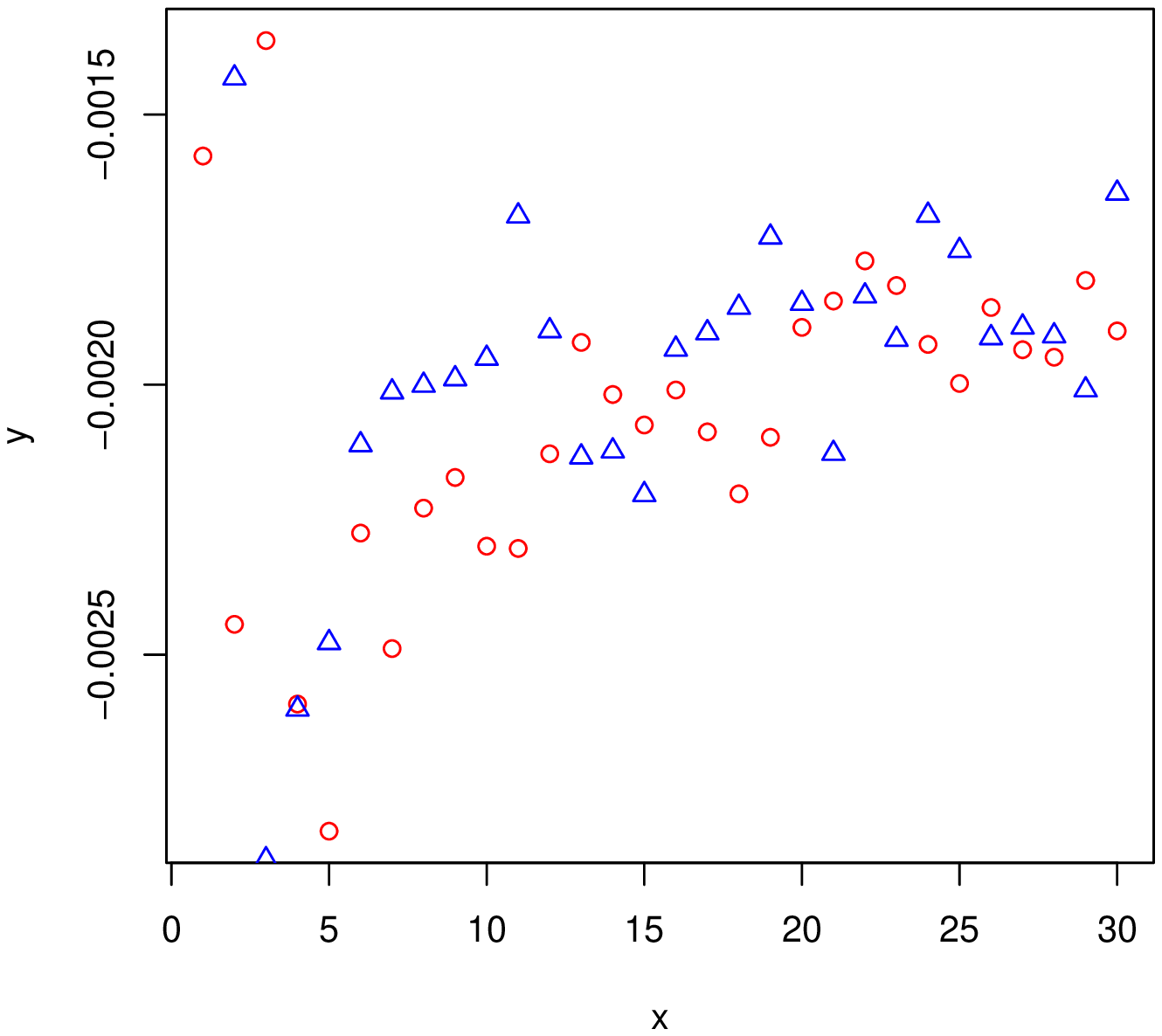}}
  \caption{$x$-axis: $K$; $y$-axis: $\zeta_2$. Circles: raw HF; triangles: HF regressed over Mkt-RF and Fama-French risk factors SMB, HML, WML. See Section \ref{cluster.num} for details.}\label{MedianCorr30}
\end{figure}

{}Let
\begin{eqnarray}
 &&\zeta_1 \equiv \mbox{Mean}\left({\widetilde \Phi}_{ij},~i\not=j\right)\\
 &&\zeta_2 \equiv \mbox{Median}\left({\widetilde \Phi}_{ij},~i\not=j\right)
\end{eqnarray}
By looking at $\zeta_1$ and $\zeta_2$ as a function of $K$, we can determine the value of $K$ above which $\zeta_1$ and $\zeta_2$ no longer decrease substantially. This value of $K$ estimates (the upper bound for) the number of clusters $F$ for the correlation matrix $\Psi_{ij}$. For the aforementioned case of the HF data the graphs for $\zeta_1$ and $\zeta_2$ as functions of $K$ are given by Fig.\ref{MeanCorr30} and Fig.\ref{MedianCorr30}. From these graphs we can deduce that $F$ is between 5 and 10, which is consistent with the above lower bound results based on the largest eigenvalue method. Note that the latter method is based on a binary cluster assumption, whereas the former method is based on non-binary analysis using principal components. These complementary methods can be used to estimate the number of clusters in the case of much larger $N$. Also, note that to use the principal component method, the correlation matrix must be positive-definite, so in Fig.\ref{MeanCorr30} and Fig.\ref{MedianCorr30} we deformed the correlation matrix using the method discussed in Subsection 3.1 of (Kakushadze, 2014a) based on (Rebonato and J\"ackel, 1999) with the non-positive eigenvalues replaced by the smallest positive eigenvalue (see footnote \ref{foot.RJ}).

\section{Is There a Limit to Turnover Reduction?}\label{theorem}

{}The answer is yes -- assuming the number of underlying tradable instruments is finite. Indeed, in this case the investment level is finite, and if we assume that by adding more and more alphas, {\em i.e.}, by taking $N \rightarrow \infty$, we can reduce the turnover indefinitely, we would arrive at a {\em static} portfolio of finite underlying tradable instruments. A dollar-neutral (or, more generally, ``market-neutral") static portfolio cannot have high Sharpe and return. Long-only static portfolios with positive returns do exist -- S\&P is an example, albeit its Sharpe ratio is low.\footnote{\, More precisely, even S\&P is not static, it is quasi-static due to periodic re-balancing, but such effects are not important for our discussion here. Also, it is not important here whether a return is positive with respect to a non-negative benchmark.} However, a dollar-neutral static portfolio with high Sharpe and return cannot persist for too long. Suppose it does. Let it have $I$ long and $I$ short dollar positions. Then by adding $I$ dollar long S\&P, we can synthesize a net $I$ dollar long portfolio with the following P\&L, volatility and Sharpe:
\begin{eqnarray}
 &&P = P_1 + P_2\\
 &&R = \sqrt{R_1^2 + R_2^2 + 2~\rho~R_1~R_2}\\
 &&S = {P\over R}
\end{eqnarray}
where $P_1$ and $P_2$ are the P\&Ls of the dollar-neutral portfolio and S\&P, $R_1$ and $R_2$ are the corresponding volatilities, and $\rho$ is the correlation between the P\&Ls, which can be assumed to be low: $\left|\rho\right| \ll 1$. Then it is evident that, if the dollar-neutral portfolio significantly outperforms S\&P, the synthetic long portfolio, which is static, outperforms S\&P. However, such a condition cannot last long as it will be quickly arbitraged away -- the synthetic portfolio is static, hence no apparent obstruction.

{}The above discussion implies that there is a limit to how much turnover can be reduced. This, in turn, implies that the number of clusters is finite so long as the number of underlying tradable instruments is finite. Once this limit is reached -- and the number of clusters for any given set of $N$ alphas can be estimated using the methods discussed in Section \ref{cluster.num} -- the development of more and more alphas no longer serves the purpose of reducing the turnover, but rather of improving the return of the portfolio by i) recycling away the alphas with degraded performance and ii) effectively enlarging the weight space and thereby improving the alpha weight optimization. Since the number of clusters is finite, they should be determined and used in order to optimize the search for new alphas to maximize the impact of i) and ii) above. And the large $N$ limit -- as in theoretical physics ('t Hooft, 1974) -- proves to be a powerful tool for understanding turnover reduction and other aspects of portfolios consisting of a large number $N$ of alphas.

\section{Comments}

{}We end this note with a few clarifying comments. First, (Kakushadze and Liew, 2014), (Kakushadze, 2014a) and this paper assume that all alphas are traded on the same execution platform. In practice this means that the $N$ alphas $\alpha_i$ are combined with some weights $w_i$ (typically, via optimization (Kakushadze, 2014c)) into a single alpha, and it is this combined alpha that is traded. This is the efficient way; the internal crossing is automatic. Second, as discussed in more detail in (Kakushadze and Liew, 2014) and (Kakushadze, 2014a), generally, it is no easy feat to precisely describe internal crossing and turnover reduction. In a portfolio consisting of a large number of underlying tradable instruments ({\em e.g.}, stocks), precise details of internal crossing depend on the detailed portfolio position and trade data. Modeling turnover reduction via alpha correlations is exactly that -- {\em modeling}. {\em E.g.}, correlations between trades and correlations between positions are not the same. However, as argued in (Kakushadze, 2014a), notwithstanding the caveats, the spectral model is expected to be a good approximation in the large $N$ limit for reasonably distributed individual turnovers and weights. In this regard, alphas being optimized, {\em i.e.}, combined into a single alpha, makes for a significant simplification -- this avoids the issue of timing trades between different alphas, which is what would have to be done had the alphas actually been traded individually. Furthermore, in the large $N$ limit the leading contribution (in the $1/N$ expansion) into the turnover reduction coefficient has a correlation-like structure (Kakushadze and Liew, 2014), and, as was argued in (Kakushadze, 2014a), this leading contribution is well-approximated (under the aforementioned conditions) by the contribution from the first principal component of the alpha correlation matrix. Third, as mentioned above, the use of the factor model approach and (binary) clusters is a convenient computational tool, which allows us to gain {\em intuitive} insight into turnover reduction beyond and complementary to the principal component approach of (Kakushadze, 2014a).

\appendix
\section{Properties of Eigenvectors}

{}Here we elaborate on (\ref{sumV}) we used in Subsection \ref{non.diag.fac.cov}, which may not be evident. To do this, instead of dealing with general factor correlation matrix ${\widehat \Psi}_{AB}$, here we will study a simpler case where
\begin{equation}
 {\widehat \Psi}_{AB} \equiv \rho,~~~A \not= B
\end{equation}
Let
\begin{equation}
 {\widetilde \chi}^{(A)} \equiv \sum_{B=1}^F \sqrt{N_B}~\chi^{(A)}_B
\end{equation}
Then from (\ref{eigen}) we have:
\begin{equation}\label{chi}
 \chi^{(A)}_C = \rho~{\widetilde \chi}^{(A)}~{\sqrt{N_C}\over{{\widehat \psi}^{(A)} - (1-\rho)~N_C}}
 \end{equation}
Multiplying both sides by $\sqrt{N_C}$ and summing over $C=1,\dots,F$, we obtain:
\begin{equation}\label{roots}
 \rho~\sum_{C=1}^F {N_C\over{{\widehat \psi}^{(A)} - (1-\rho)~N_C}} = 1
\end{equation}
The eigenvalues ${\widehat \psi}^{(A)}$ are the $F$ roots of this equation (assuming for the sake of simplicity that all $N_C$ are distinct). It further follows from (\ref{roots}) that
\begin{equation}\label{phistar}
 \rho~\sum_{C=1}^F {1\over{{\widehat \psi}^{(A)} - (1-\rho)~N_C}} = {\phi^*\over {\widehat \psi}^{(A)}}
\end{equation}
where
\begin{equation}\label{phistar1}
 \phi^* \equiv 1 + \rho~(F - 1)
\end{equation}
is the non-degenerate eigenvalue of ${\widehat \Psi}_{AB}$ -- the other $(F-1)$ eigenvalues $\phi^\prime \equiv 1 -\rho$.

{}For $A\not=B$ we have
\begin{eqnarray}
 &&\sum_{C=1}^F \chi^{(A)}_C~\chi^{(B)}_C = \rho^2~
 {\widetilde \chi}^{(A)}~{\widetilde \chi}^{(B)}~\sum_{C=1}^F {N_C \over {\left[ {\widehat \psi}^{(A)} - (1-\rho)~N_C\right]~\left[{\widehat \psi}^{(B)} - (1-\rho)~N_C\right]}} = \nonumber\\
 && = {\rho^2~{\widetilde \chi}^{(A)}~{\widetilde \chi}^{(B)} \over{(1-\rho)~\left({\widehat \psi}^{(B)} - {\widehat \psi}^{(A)}\right)}}~\sum_{C=1}^F
 \left[{{\widehat \psi}^{(A)} \over {{\widehat \psi}^{(A)} - (1-\rho)~N_C}} - {{\widehat \psi}^{(B)} \over {{\widehat \psi}^{(B)} - (1-\rho)~N_C}}\right] =  \nonumber\\
 && = 0
\end{eqnarray}
So different eigenvectors are orthogonal to each other, as they should be.

{}From (\ref{norm}) we have:
\begin{equation}\label{norm1}
 1 = \sum_{C=1}^F \left(\chi^{(A)}_C\right)^2 = \rho^2~
 \left({\widetilde \chi}^{(A)}\right)^2~\sum_{C=1}^F {N_C \over {\left[ {\widehat \psi}^{(A)} - (1-\rho)~N_C\right]^2}}
\end{equation}
The last sum can be tackled as follows. Let $\partial$ denote the derivative w.r.t. $\rho$, {\em e.g.}:
\begin{equation}
 \partial {\widehat \psi}^{(A)} \equiv {\partial {\widehat \psi}^{(A)} \over {\partial \rho}}
\end{equation}
Differentiating (\ref{phistar}) w.r.t. $\rho$ and rearranging terms, we obtain:
\begin{equation}
 \rho^2~
 \left[{\widehat \psi}^{(A)} + (1-\rho)~\partial {\widehat \psi}^{(A)}\right]~\sum_{C=1}^F {N_C \over {\left[ {\widehat \psi}^{(A)} - (1-\rho)~N_C\right]^2}} =
 \phi^* - \rho~\partial \phi^*
\end{equation}
Together with (\ref{phistar1}) and (\ref{norm1}) this gives the following simple expression:
\begin{equation}
 \left({\widetilde \chi}^{(A)}\right)^2 = {\widehat \psi}^{(A)} + (1-\rho)~\partial {\widehat \psi}^{(A)}
\end{equation}
For our purposes here we will not need the explicit form of $\partial {\widehat \psi}^{(A)}$.

{}We are interested in understanding the behavior of
\begin{equation}
 \left|\sum_{i=1}^N V^*_i \right| = \left|{\widetilde \chi}^*\right|
\end{equation}
where ${\widetilde \chi}^*\equiv {\widetilde \chi}^{(A)}$ for the value of $A$ for which ${\widehat \psi}^{(A)} = {\widehat \psi}^*$. The largest eigenvalue monotonically increases as $\rho$ increases from 0 to 1, {\em i.e.}, $\partial{\widehat \psi}^* > 0$. This then implies (\ref{sumV}) as we have (\ref{maxpsi}). Also, note that $\det({\widehat \Psi}^\prime) = \det({\widehat \Psi}) ~\prod_{A=1}^F N_A$, which implies that as $\rho$ approaches 1, $(F-1)$ eigenvalues go to zero, while the largest eigenvalue goes to $N$ -- recall that ${\widehat \Psi}$ has one eigenvalue equal $\phi^* = 1 + (F-1)~\rho$ and $(F-1)$ eigenvalues equal $\phi^\prime = 1-\rho$.

\begin{figure}
\centerline{\epsfxsize 4.truein \epsfysize 4.truein\epsfbox{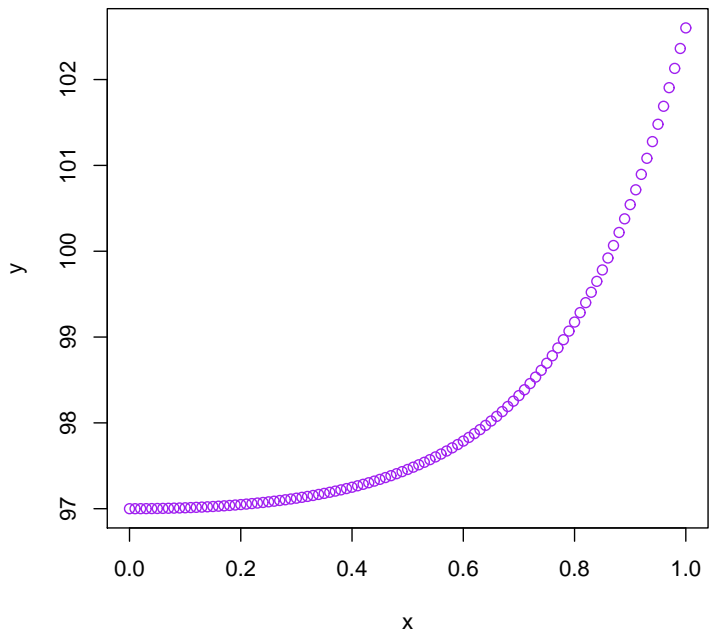}}
\caption{$x$-axis: $\rho$; $y$-axis: ${\widehat \psi}^*$. See Appendix A for details.}
\label{psigraph}
\end{figure}

{}Fig.\ref{psigraph} is a graph of ${\widehat \psi}^*$ {\em vs.} $\rho$ for a randomly constructed matrix with $F = 50$ and $N = 2061$. Also, $F=2$ provides analytical insight into the eigenvalue structure. Let $N_1 > N_2$. The eigenvalues are given by
\begin{eqnarray}
 &&{\widetilde \psi}^{(1)} = {1\over 2}~\left[N_1 + N_2 + \sqrt{\left(N_1 - N_2\right)^2 + 4~N_1~N_2~\rho^2}\right]\\
 &&{\widetilde \psi}^{(2)} = {1\over 2}~\left[N_1 + N_2 - \sqrt{\left(N_1 - N_2\right)^2 + 4~N_1~N_2~\rho^2}\right]
\end{eqnarray}
For $\rho=0$ the eigenvalues are $N_1$ and $N_2$. The larger eigenvalue monotonically increases to $N = N_1 + N_2$ as $\rho$ increases from 0 to 1, while the lower eigenvalue decreases to 0.

\section{Non-binary Case}

{}Here our goal is to estimate (\ref{Vstar.1}). To do this, let (here we use the notations of Subsection 5.4)
\begin{eqnarray}
 &&\Lambda_{iA} \equiv \lambda_A + {\widetilde \Lambda}_{iA}\\
 &&\sum_{i=1}^N {\widetilde\Lambda}_{iA}\equiv 0
\end{eqnarray}
{\em I.e.}, ${\widetilde\Lambda}_{iA}$ are obtained from $\Lambda_{iA}$ by demeaning its columns.  We have
\begin{eqnarray}
 &&\zeta_A\equiv \sum_{i=1}^N V^{(A)}_i = {N\over\sqrt{\psi^{(A)}}} \sum_{B=1}^F \lambda_B~W_{BA}\\
 &&Q_{AB} = N~\lambda_A\lambda_B + {\widetilde Q}_{AB}\\
 &&{\widetilde Q}_{AB} \equiv \sum_{i=1}^N {\widetilde\Lambda}_{iA}~{\widetilde\Lambda}_{iB}
\end{eqnarray}
Let
\begin{eqnarray}
 &&{\widetilde Q} = {\widetilde W}~{\widetilde Z}~{\widetilde W}^T\\
 &&{\widetilde Z} = \mbox{diag}\left(q_A\right)
\end{eqnarray}
where ${\widetilde W}$ is the $F\times F$ matrix, normalized such that ${\widetilde W}^T~{\widetilde W} = 1$, whose columns are the right eigenvectors of ${\widetilde Q}_{AB}$ corresponding to the eigenvalues $q_A$. Note that, up to a similarity transformation, $Q_{AB}$ is given by a one-factor model:
\begin{eqnarray}
 && Q = {\widetilde W} ~{\widehat Q} ~{\widetilde W}^T\\
 &&{\widehat Q} \equiv q_A~\delta_{AB} + {\widehat \lambda}_A~{\widehat \lambda}_B\\
 &&{\widehat \lambda}_A \equiv \sqrt{N}\sum_{B=1}^F \lambda_B~{\widetilde W}_{BA}
\end{eqnarray}
Instead of dealing with the most general case, here we will assume that the eigenvalues $q_A$ are identical: $q_A\equiv q$. This can be thought of as an approximation where volatilities $q_A$ in the above one-factor model are replaced (be it directly or logarithmically), {\em e.g.}, by mean (or median) volatility.\footnote{\, This is analogous to our simplifying approximation in Subsection 5.1, where we assumed uniform specific risk across each cluster.} This approximation will suffice for our purposes here simplifying the math to the point where it is illuminating. We then have
\begin{eqnarray}
 &&Q_{AB} = N~\lambda_A\lambda_B + q~\delta_{AB}\\
 &&\psi^* = N~\chi^2 + q\\
 &&\chi \equiv \sqrt{\sum_{A=1}^F \lambda_A^2}\\
 &&\psi^\prime = q\\
 &&W_{A*} = {\lambda_A\over\chi}
\end{eqnarray}
Here $\psi^*$ is the largest eigenvalue of $Q_{AB}$, while $\psi^\prime$ are the other (degenerate) $F-1$ eigenvalues. Also, $W_{A*}$ is the right eigenvector of $Q_{AB}$ corresponding to $\psi^*$. Thus:
\begin{equation}\label{zeta.star}
 \zeta_* = {N\over\sqrt{\psi^*}} \sum_{B=1}^F \lambda_B~W_{B*} = {N~\chi\over \sqrt{\psi^*}}
\end{equation}
Here we need to understand the possible values of $\chi$ and $q$. First, note that $\chi$ cannot vanish. Indeed, if it does, then all $\lambda_A \equiv 0$, which would then imply that $\sum_{j=1}^N \Psi_{ij} \equiv 0$, $i=1,\dots,N$. This would imply that at least some off-diagonal elements of $\Psi_{ij}$ are negative. Let $\Psi_{i_1 j_1} < 0$ for some $i_1 \neq j_1$. Then, by flipping the sign of the alpha corresponding to $j_1$, $\alpha^\prime_{j_1}\equiv -\alpha_{j_1}$ (and keeping all other alphas unchanged, $\alpha^\prime_j \equiv \alpha_j$, $j\neq j_1$), we would get the corresponding correlation matrix $\Psi^\prime_{ij}$  such that $\sum_{j=1}^N \Psi^\prime_{i_1 j} = -2~\Psi_{i_1 j_1} > 0$. So, $\chi > 0$. Also, recall that $\mbox{Tr}(Q) = N$, which gives $N~\chi^2 + F~q = N$, so $q < N/F$ and $\psi^* = N~\chi^2 + q = N - (F-1)~q > N/F$, which is consistent with (\ref{maxpsi.1}). However, there is a stronger bound on $\chi$. A simple argument goes as follows. Note that $\sum_{i,j=1}^N \Psi_{ij} = N^2\sum_{A=1}^N \lambda_A^2 = N^2\chi^2$. On the other hand, $\sum_{i,j=1}^N \Psi_{ij} = N^2\rho^\prime = N(1 +(N-1){\overline \rho})$, where ${\overline\rho}$ is the mean correlation (see footnote \ref{rho.prime}). On general grounds (Kakushadze, 2014a), we expect that $\rho_* = \gamma~\rho^\prime$, where $\gamma\sim 1$ (and typically $\gamma > 1$). It then follows that $\chi\approx \sqrt{\rho_*/\gamma}$ and since $\rho_* = \psi^*\zeta_* / N^{3/2}$, where $\zeta_*$ is given by (\ref{zeta.star}), we get
\begin{equation}
 \rho_* \approx {\psi^* \over\gamma N}\gsim {1\over\gamma F}
\end{equation}
So, in the non-binary case we get a higher bound for $\rho_*$ than in the binary case. This is because in the binary case we actually have diagonal $Q_{AB} = N_A~\delta_{AB}$ ($N_A$ is defined in Section 5), and $\zeta_A = \sqrt{N_A}$ in that case (for a diagonal factor covariance matrix).

\newpage

\end{document}